\documentclass[aps,prb,twocolumn,superscriptaddress]{revtex4-1}

\usepackage{tabularx}

\usepackage[version=3]{mhchem} 
\usepackage{graphicx}
\usepackage{xcolor}
\usepackage{textcomp}

\usepackage[flushleft]{threeparttable}
\usepackage{booktabs,multirow}

\usepackage[breaklinks]{hyperref}
\hypersetup{
	pdfnewwindow=true,      
	colorlinks=true,       
	linkcolor=blue,          
	citecolor=blue,        
	filecolor=blue,      
	urlcolor=blue           
}

\begin{document}

\author{Viktoria V. Ivanovskaya}
\affiliation{Dipartimento di Scienze Chimiche, Università degli Studi di Padova, 35131 Padova, Italy }
\author{Alberto Zobelli}
\affiliation{Universit\'e Paris-Saclay, CNRS, Laboratoire de Physique des Solides, 91405, Orsay, France}
\author{Andrea Basagni}
\affiliation{Dipartimento di Scienze Chimiche, Università degli Studi di Padova, 35131 Padova, Italy }
\author{Stefano Casalini}
\affiliation{Dipartimento di Scienze Chimiche, Università degli Studi di Padova, 35131 Padova, Italy }
\author{Luciano Colazzo}
\affiliation{Center for Quantum Nanoscience, Institute for Basic Science (IBS), Seoul 03760, Republic of Korea; Ewha Womans University, Seoul 03760, Republic of Korea}
\author{Francesco de Boni}
\affiliation{Dipartimento di Scienze Chimiche, Università degli Studi di Padova, 35131 Padova, Italy }
\author{Dimas G. de Oteyza}
\affiliation{Nanomaterials and Nanotechnology Research Center (CINN), CSIC-UNIOVI-PA, 33940 El Entrego, Spain}
\author{Mauro Sambi}
\affiliation{Dipartimento di Scienze Chimiche, Università degli Studi di Padova, 35131 Padova, Italy }
\author{Francesco Sedona}
\email{francesco.sedona@unipd.it}
\affiliation{Dipartimento di Scienze Chimiche, Università degli Studi di Padova, 35131 Padova, Italy }

\title[]{On-surface synthesis and evolution of self-assembled poly($p$-phenylene) chains on Ag(111): a joint experimental and theoretical study}

\begin{abstract}
	The growth of controlled 1D carbon-based nanostructures on metal surfaces is a multistep process whose path, activation energies and intermediate metastable states strongly depend on the employed substrate. Whereas this process has been extensively studied on gold, less work has been dedicated to silver surfaces, which have a rather different catalytic activity. In this work, we present an experimental and theoretical investigation of the growth of poly-$p$-phenylene (PPP) chains and subsequent narrow graphene ribbons starting from 4,4''-dibromo-$p$-terphenyl molecular precursors deposited at the silver surface. By combing scanning tunneling microscopy (STM) imaging and density functional theory (DFT) simulations, we describe the molecular morphology and organization at different steps of the growth process and we discuss the stability and conversion of the encountered species on the basis of calculated thermodynamic quantities. Unlike the case of gold, at the debromination step we observe the appearance of organometallic molecules and chains, which can be explained by their negative formation energy in the presence of a silver adatom reservoir. At the dehydrogenation temperature the persistence of intercalated Br atoms hinders the formation of well-structured graphene ribbons, which are instead observed on gold, leading only to a partial lateral coupling of the PPP chains. We numerically derive very different activation energies for Br desorption from the Ag and Au surfaces, thereby confirming the importance of this process in defining the kinetics of the formation of molecular chains and graphene ribbons on different metal surfaces.
\end{abstract}

\maketitle

\section{Introduction}
In the last 20 years much attention has been devoted to the production and study of intrinsic properties of new building blocks for future electronics such as 2D layered materials or molecular semiconductors. The high atomic level control required by these new architectures can hardly be achieved by top-down approaches conventionally employed in current nanoelectronic fabrication. A very promising and effective strategy, generally called ``on-surface synthesis'' (OSS), is based on a bottom-up approach where supramolecular arrangements and nanostructures can be built up with atomic precision from well-designed molecular precursors self-organizing and reacting at metallic surfaces.\cite{Zhou2020,Chen2020,Held2017,Goronzy2018,Han2021,Clair2019} 

In this scenario, the most successful results have been obtained in the synthesis of long 1D and 2D structures starting from precursors with two functional groups. The examples are many, starting from the pioneering works of Leonhard Grill,\cite{Lafferentz2009} passing through the extensive study of graphene nanoribbons,\cite{Houtsma2021} up to the attempts to obtain graphyne-like molecular chains with the alternation of sp and sp$^2$ hybridized carbon atoms.\cite{Li2019} Most of these results have been obtained through a multistep protocol\cite{Dong2015,Eichhorn2014} exploiting organic monomers with halogen functions deposited on a metal. According to an Ullmann-like reaction, C-halogen bonds are thermally cleaved and halogen atoms are adsorbed at the surface to obtain surface-stabilized biradical species, which later merge in linear polymer chains. A successive thermal activation step is required for intra or extra-molecular dehydrogenative cyclization to produce more complex structures such as graphene nanoribbons (GNRs).\cite{Houtsma2021} 
Key parameters of the synthesis process can be derived from the in-depth knowledge of adsorption energies and activation barriers for molecular migration, bond breaking and recombination. While these quantities might be hardly accessed experimentally, numerical simulations have provided accurate estimations for a large set of molecular structures at different metal surfaces.\cite{hofmann2021first,draxl2014organic}

OSS has been efficiently employed to produce high-quality extended  poly-$p$-phenylene (PPP)  chains, which can be merged into narrow GNRs.\cite{Merino-Diez2017} PPP is a $\pi$-conjugated  large band-gap polymer formed by a line of phenyl rings linked in  the para position, which itself can be considered as the simplest  armchair graphene nanoribbon (AGNR) with N=3.\cite{Chen2020} PPP molecules have also been proposed for possible optoelectronic devices such as blue-emitting diodes.\cite{Grimsdale2009,Baur1998,Shacklette1980,Grem1995}
The main steps of the PPP growth protocol are similar for different substrates but significant differences have been observed in activation energies and growth paths.\cite{Fritton2019,Grill2020} For instance, the formation of intermediate organometallic chains has been identified on silver and copper surfaces\cite{Zhou2017,Li2021,Mannix2015,Galeotti2019,Cirera2017,Vasseur2016}  while it has never been observed on gold.\cite{Basagni2015,DeBoni2020,Basagni2016,Stolz2020,Abadia2017,Piquero-Zulaica2018,Merino-Diez2018} Furthermore, the debromination can be reversed on gold but this process is hindered on silver by the higher stability of individual adsorbed Br atoms and the possible trapping of metal adatoms at the radical site.\cite{Stolz2020} The molecular arrangement of the PPP molecular chains is also expected to change as a function of the substrates due to different molecule-metal interactions but this information can hardly be accessed experimentally. Whereas PPP chains have been extensively studied on gold substrates, fewer works have been dedicated to silver substrates.\cite{Chung2012,Lu2020,Dai2019} The differences observed at the different growth steps over various metal surfaces lead to a flawed picture which precludes a clear understanding of the role exerted by the substrate in defining the growth dynamics.

In this work we present a comprehensive study of the multistep growth of PPP chains and GNRs starting from 4,4''-dibromo-$p$-terphenyl (DBTP) molecular precursors adsorbed at silver surfaces. Long annealing times were employed at each growth step to ensure that a steady state was reached under the given conditions. The structure and organization of all the obtained molecular complexes were investigated by scanning tunneling microscopy (STM), low energy electron diffraction (LEED), X-ray photoelectron spectroscopy (XPS) and near edge X-ray absorption fine structure (NEXAFS) measurements. Complemental simulations conducted in the framework of the density functional theory (DFT) have been performed to give a precise description at the atomic level of the supramolecular organization and molecular reshaping at all steps of the growth process. In comparison with the case of gold, we observe the appearance of metastable organometallic complexes (individual molecules and chains) already at room temperature. The difference between the two substrates is here explained on the basis of the calculated formation energy of organometallic complexes in the presence of metal adatoms reservoirs. At higher temperatures, residual Br atoms intercalating the chains hinders the formation of well-structured GNRs, which are instead observed on gold, leading only to a partial lateral coupling of the PPP chains. A comparative theoretical analysis of H and Br desorption paths and activation energies on gold and silver stresses the importance of these processes in defining the kinetics of the formation of molecular chains and graphene ribbons.

\section{Methods}

\paragraph*{Sample Preparation}
All experiments have been performed at a base pressure of $ 8 \times 10^{-10}$  mbar. The Ag(111) crystal was cleaned by repeated cycles of 1.5 keV \ce{Ar+} sputtering and annealing at 670 K until a clean surface with sufficiently large terraces was obtained as confirmed by STM imaging. Commercially available DBTP molecules were deposited from a pyrolytic boron nitride crucible held at $\sim 390$ K. All the reported experiments have been performed starting with a high coverage ranging between 0.8 and 0.9 ML as calibrated by the C 1s XPS signal; we define one monolayer (1ML) as the surface fully covered by the adsorbed DBTP molecules forming the [4 2, 2 6] superstructure (see Results and Discussion). During deposition the surface was always held at room temperature (RT) and the polymerization was activated by subsequent thermal annealing. The  annealing temperature was kept for at least 5 h to allow the system to evolve until it reached a stationary state under the given conditions; the samples were then cooled to RT and analyzed.

\paragraph*{STM Imaging}
Experiments were performed with an Omicron scanning tunneling microscope (VT-STM). All STM measurements were carried out at RT in constant-current mode using an electrochemically etched Pt-Ir tip. The STM data were processed with the WSxM software.\cite{Horcas2007} Moderate filtering was applied for noise reduction.

\paragraph*{LEED}
The diffraction patterns have been performed with a OCI LPS075D rear view instrument and interpreted with the LEEDpat software.

\paragraph*{XPS}
Measurements were performed in situ at RT using a VG Scienta XM 650 X-ray source. The X-rays produced were monochromatized using a VG Scienta XM 780 monochromator optimized for the Al K$\alpha$ line (1486.7 eV). Photoelectrons were collected and analyzed with a Scienta SES 100 electron analyzer fitted to the STM preparation chamber. The fitting of XPS peaks has been done with the  XPSPeak software.

\paragraph*{NEXAFS}
Measurements were performed at the ALOISA beamline\cite{Floreano1999} of the ELETTRA synchrotron (Trieste). The C K-edge NEXAFS spectra were measured via the partial electron yield with a channeltron and a negatively biased (230 V) grid in front of it to reject low energy secondary electrons. The spectra have been energy-calibrated a posteriori by the characteristic absorption signal of the carbon K-edge in the I0 signal (drain current on the last mirror).

\paragraph*{DFT}
All calculations have been performed using the OpenMX code.\cite{Ozaki2003,OpenMXsite} The exchange-correlation potential was expressed in the generalized gradient approximation using the Perdew-Burke-Ernzerhof (PBE)\cite{Perdew1996} formalism.  Dispersion interactions have been included using the Grimme-D2 method\cite{Grimme2006} which provides a good agreement with the experimentally observed PPP molecule reconformation on Ag and Au surfaces. In comparison, the Grimme-D3 method does not satisfactorily reproduce the flattening of the PPP chains observed at silver surfaces, while it gives similar results for the Au surface.
We used norm conserving fully relativistic pseudopotentials including a partial core correction and a basis set of optimized numerical pseudoatomic orbitals. Charge densities and potentials were determined on a real-space grid with a mesh cutoff energy of 210 Ry. The systems have been fully relaxed and a $6\times 6 \times 1$ k-point mesh has been proved to provide a sufficient sampling of the Brillouin zone. 
Supercell lattice parameters have been derived from the DFT-PBE optimized values for the primitive cubic bulk phase, which is 4.17 \AA \- for both gold and silver. We used a slab thickness of 14 metal layer,s which grants a correct description of the surface electronic states.\cite{Motornyi2018} A 40 \AA \- vacuum separation between the slabs has been employed to minimize interaction along the c direction.
Simulated constant current scanning tunneling microscopy (STM) images have been obtained within the Tersoff-Hamann approximation\cite{Tersoff1985} where the tunneling current is proportional to the local density of surface states at the tip position integrated from the applied voltage bias to the Fermi level. For comparison with experiments, images have been later convoluted with a Gaussian function to reproduce tip size effects. Transition states have been calculated using the nudged elastic band method as implemented using a total number of 20 images.

\section{Results and Discussion}

\subsection{Structure}
\begin{figure*}
	\includegraphics[width=\textwidth]{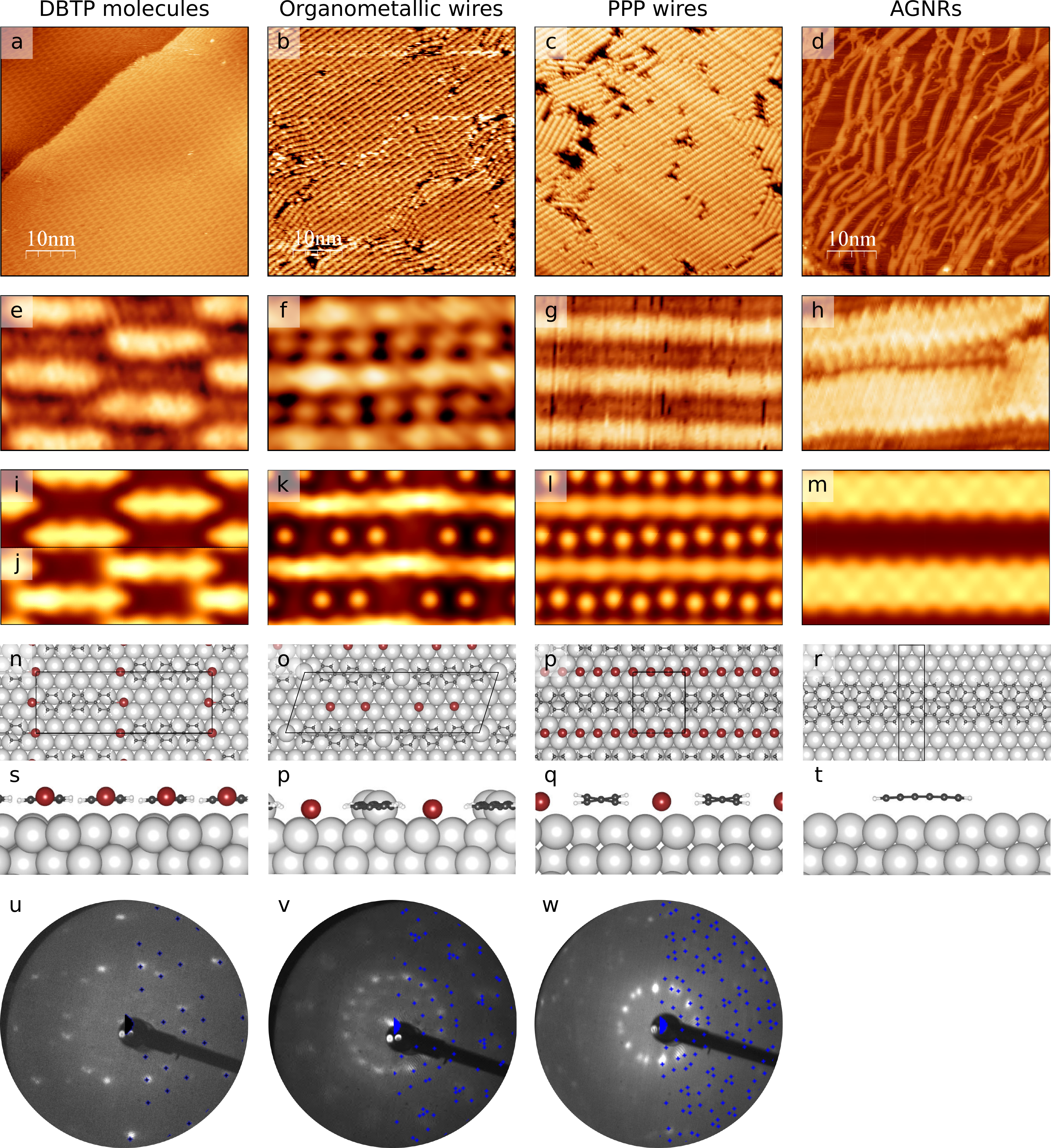}
	\caption{(a-d) STM large scale ($50\times 50$ nm) and (e-h) high resolution ($1.3 \times 2$ nm) images of the principal surface nanostructures obtained by thermal annealing and detailed in the text. (i-m) Respective simulated STM images; (i) adsorbed DBTP molecules and (j) one end Ag substituted DBTP molecules. (n-r) Top and (s, t) side views of DFT optimized structural models where lines indicate the surface supercell. (u-w) Experimental LEED pattern and the associated simulation based on the reported unit cells. STM parameters: a) V=-0.9 V, I= 5.0 nA; b) V=1.0, V I=2.8 nA; c) V=-0.9, V I= 1 nA; d) V=1.0, V I=9.5 nA; e) V=0.1 V, I= 4.8 nA; f) V=-0.3 V, I= 4.0 nA; g) V=-1.0 V, I=1.3 nA; h) V= -0.9 V, I= 4.0 nA.}
	\label{Fig:STM}
\end{figure*}

\begin{figure*}
	\centering
	\includegraphics[width=0.85\textwidth]{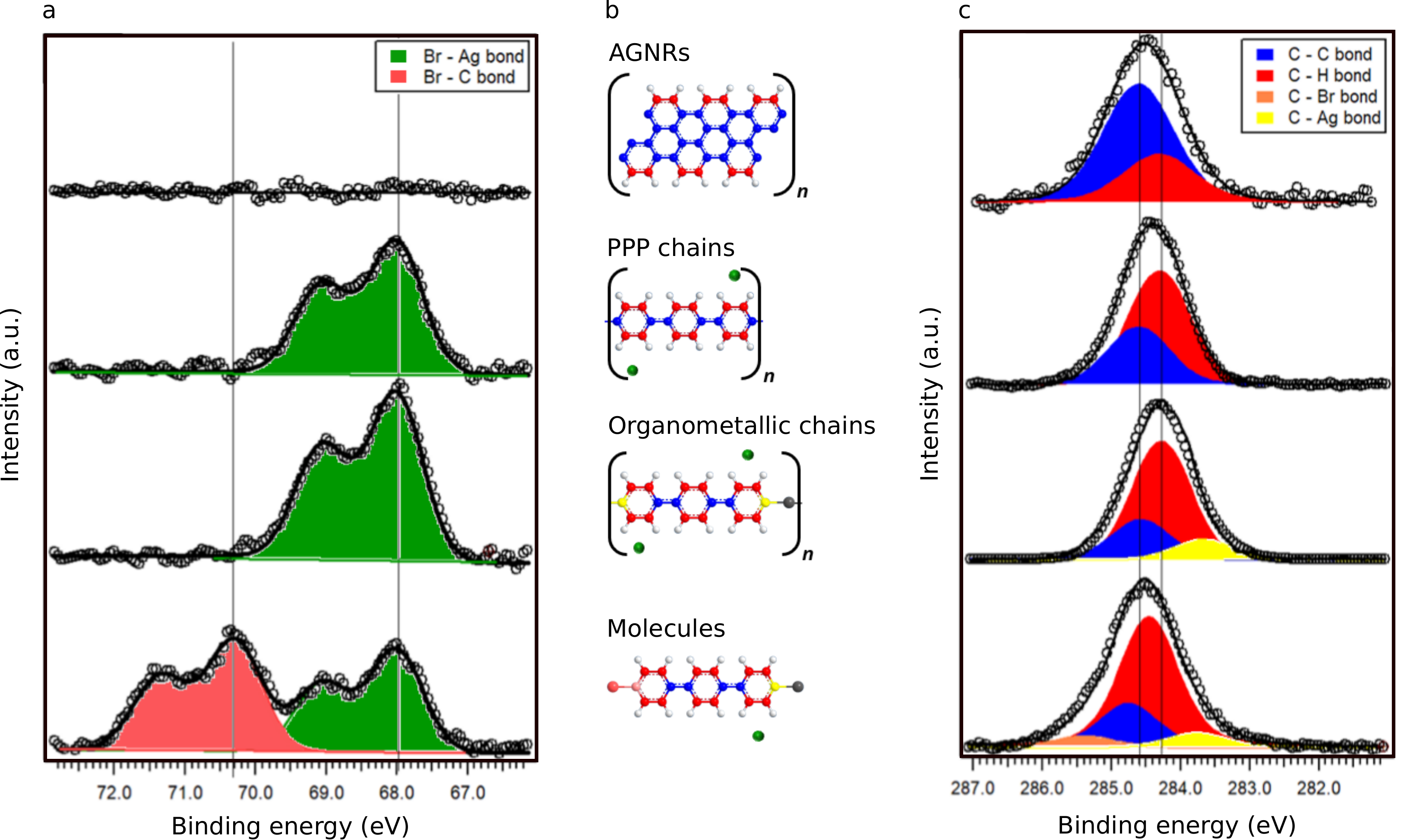}
	\caption{(a) C 1s and (c) Br 3d XPS spectra and fitting components of different molecular structures adsorbed on Ag (111). From top to bottom: AGNRs, PPP chains, organometallic chains and molecules. (b) Corresponding structural models where the atom color corresponds to the different bonding types identified in the XPS spectra.}
	\label{Fig:XPS}
\end{figure*}

In Figure \ref{Fig:STM} we show a series of STM and LEED images illustrating the different stages of the formation of PPP chains and successively GNRs at the Ag(111) surface. Experimental observations, especially LEED patterns, give a clear view of long-range order but they do not provide precise information on the reshaping of the molecules induced by their interaction with the substrate. Therefore, to complement the experimental information, in the same figure we  report also structural models relaxed by DFT calculations and the corresponding STM image simulations. In Figure \ref{Fig:XPS}  we finally present XPS spectra acquired for the C 1s and the Br 3d peaks.

\paragraph{DBTP Molecules}
In the first step, the DBTP molecular precursor is adsorbed at RT onto the metal surface, reaching a surface coverage of about 0.8-0.9 ML. The molecules organize in a chessboard structure commensurate with the substrate (Figure  \ref{Fig:STM}a,e), which, can be attributed to a [4 2 , 2 6] supercell (Figure 1.n) on the basis of the observed LEED pattern (Figure \ref{Fig:STM}u). A similar organization has been identified for DBTP molecules on the Au(111) surface after mild annealing at 320 K.\cite{Basagni2016} The experimental STM images show that a fraction of the molecules have a different local electron density at the two ends. STM image simulations obtained considering a partial substitution of Br atoms with Ag can perfectly reproduce the recorded images (Figure \ref{Fig:STM}i (\ref{Fig:STM}j)) showing the simulated image of the pristine (Ag-substituted) DBTP molecules. The presence of free Br atoms has been confirmed by XPS measurements conducted just after the deposition. The Br 3d XPS peak (Figure \ref{Fig:XPS} a bottom panel) shows two spin-orbit doublets with a similar integrated intensity. The first doublet, centered at 68 eV, can be attributed to the 3d$_{5/2}$ states of Br atoms adsorbed at the surface, while the second, centered at 70.3 eV, can be attributed to the 3d$_{5/2}$ states of Br atoms still bonded to the molecules. These observations indicate that debromination and the subsequent formation of organometallic complexes through the capture of Ag adatoms can occur already at RT as a consequence of the well-documented higher reactivity of silver with respect to gold, where the molecules remain intact at RT.\cite{Basagni2016} The corresponding C1s peak centered at 284.5 eV can be fitted by assuming the presence of four types of bonds: C-C, C-H, C-Br and C-Ag and the area of their respective contributions is in line with the corresponding atomic fraction highlighted in the model reported in Figure \ref{Fig:XPS}b. 

\paragraph{Organometallic Chains}
After keeping the sample for 48 h at RT or after a few hours of mild annealing at 320 K, we observe that the DBTP molecular organization at the metal surface transforms in a pattern of chains characterized by regular bright bumps (Figure \ref{Fig:STM}b,f). These structures can be attributed to organometallic chains formed by terphenyl molecules linked by the Ag adatoms naturally present at the surface (Figure \ref{Fig:STM}o). This result is further confirmed by XPS measurements that now show a single Br 3d peak related to the sole presence of individual Br atoms adsorbed at the surface (Figure \ref{Fig:XPS}.a). The C 1s peak is also slightly downshifted with respect to the as-deposited molecules towards 284.3 eV, indicating the increase of C-Ag bonds (Figure  \ref{Fig:XPS}c). The here observed evolution of the DBTP molecules at the Ag surface strongly differs from that observed on Au surfaces, for which the formation of organometallic chains has not been reported.\cite{Basagni2015}

The recorded complex LEED pattern (Figure \ref{Fig:STM}v) corresponds to the formation of a [4 1, 0 11] supercell. In the related atomic model, the metal atoms linking the debrominated DBTP molecules lie in two inequivalent surface fcc and hcp hollow sites (Figure \ref{Fig:STM}o). These metal atoms pin the molecules to the surface resulting, after structural optimization, in a maximum distance from the surface of 2.8 {\AA} for the carbon atoms at the molecule axis. The simulated constant current STM images (Figure \ref{Fig:STM}k) excellently match the experimental data reported in Figure \ref{Fig:STM}f, which validates the proposed structural model and the details of the relaxed structure. As in the case of partially debrominated molecules, the bright spots within the chains can be easily identified as interlinking metal atoms. The bright spots between the chains correspond instead to Br adatoms.

\paragraph{PPP Chains}
Further thermal annealing at 400 K leads to the complete ejection of the metal atoms from the organometallic chains and the subsequent formation of PPP chains. The surface is now characterized by long chains separated by Br atoms adsorbed at the surface, as reported in Figure \ref{Fig:STM}c,g. The distance between adjacent parallel chains is about 10 \AA. The long-range order is documented by the LEED pattern reported in Figure \ref{Fig:STM}w and it corresponds to a [4 1, 0 3] superstructure matrix. The model reported in Figure \ref{Fig:STM}p assumes that the organic chains are aligned along the Ag[$1\bar{1}0$] direction and that two consecutive hexagonal rings are centered on fcc and hcp hollow sites of the substrate, respectively, with almost no strain (about 1\%), complying with the experimental unit cell derived from the LEED pattern. The C 1s XPS peak shows a shift toward higher binding energy with respect to the organometallic chains due to the breaking of C-Ag bonds. Instead, the Br 3d feature does not present any change with respect to the previous growth step, indicating that at this temperature Br desorption is still not activated, see Figure  \ref{Fig:XPS}a.

The optimized equilibrium structure of the PPP chains in a molecular crystal corresponds to a nonplanar configuration where hexagonal rings are alternately twisted by a torsional angle of 20°–30°\cite{Ambrosch-Draxl1995} due to the steric hindrance of H atoms in ortho- position with respect to the phenyl-phenyl bond. Considering free PPP chains, this angle gets as high as 27.4°\cite{Ambrosch-Draxl1995} or 34.8°\cite{Miao1998} as estimated by DFT simulations. The adsorption of the molecular chains at a metal surface results in a reduction of the torsional angle, since dispersion interactions promote the flattening of hexagonal rings toward the surface (Figure \ref{Fig:STM}q). The conformation of the chains at different metal surfaces depends therefore on  competition between van der Waals interactions and the energy associated with the twist of the rings. 

The molecules torsional angles can be estimated from NEXAFS measurements by analyzing the intensity of the $\pi^*$ peak as a function of the polarization direction with respect to the normal to surface. In the case of PPP chains grown on an Au vicinal surface this angle has been estimated to be as high as 40$^\circ$.\cite{Basagni2016} In this work we have performed similar polarization-dependent NEXAFS measurements for DBTP molecules deposited on an Ag(554) vicinal surface, over which well-aligned PPP chains parallel to the vicinal steps edge direction can be obtained.

\begin{figure}[b]
	\includegraphics[width=\columnwidth]{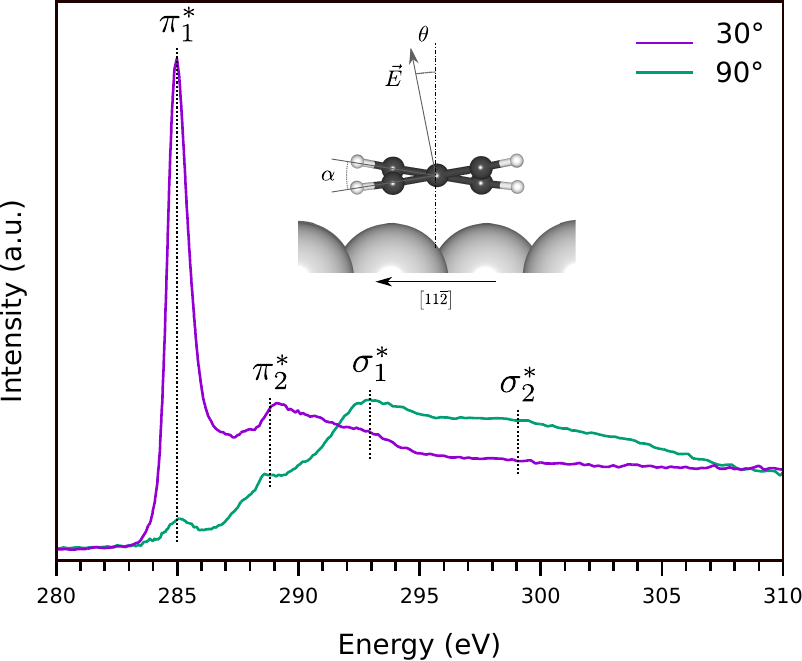}
	\caption{C K-edge NEXAFS spectra for the aligned PPP chains on Ag (554). Insert: schematic side view of the adsorbed chain structure during the NEXAFS experiment. The spectra were collected with the in-plane polarization defined by the surface normal and the [$11\bar{2}$] direction (perpendicular to the chains); $\theta$ is defined as the angle between the polarization and the surface normal, and $\alpha$ is the torsional angle between adjacent hexagonal rings.}
	\label{Fig:NEXAFS}
\end{figure}

As reported in Figure \ref{Fig:NEXAFS}, the C K-edge exhibits four peaks at 285.0 eV ($\pi_1^*$), 289.0 eV ($\pi_2^*$), 292.9 eV ($\sigma_1^*$), and a broad peak around 300 eV ($\sigma_2^*$). Comparing the angle-dependent intensity of the $\pi_1^*$ resonance and the predicted resonance intensity according to a Stöhr-derived equation for 2-fold substrate symmetry, as done for PPP on gold,\cite{Basagni2016} we derive a torsional angle between nearby benzene rings of about 24$^\circ$. NEXAFS linear dichroism hence confirms that the PPP chains on the silver surface adopt a more planar conformation as compared to the gold one, thereby indicating stronger vertical interactions. A low angle around 20$^\circ$ had been proposed on the basis of high resolution STM images.\cite{Piquero-Zulaica2018} 

This behavior was further analyzed by DFT simulations. In the case of the silver substrate, we obtain an optimized distance between the interphenyl bonding C-C atoms and the surface of 2.8 {\AA} and a torsional angle of 18$^\circ$. A gold substrate leads instead to weaker interactions and therefore to a higher distance of 3.8 {\AA} and a torsional angle of 56$^\circ$. On the other side, it has been found that stronger interactions on a copper surface\cite{Vasseur2016} lead to very short distances (2.2 {\AA}, a value very close to those of graphene adsorbed at the same metal surface) and to the formation of flat chains.

\paragraph{AGNR}
The desorption of Br atoms from the surface begins at the temperature of 650 K, as shown by the decrease of the Br XPS signal. At this temperature the cyclodehydrogenation between different PPP chains is activated, leading to the formation of armchair graphene nanoribbons (AGNRs) belonging to the 3p family. Further annealing at 700 K leads to the complete desorption of the Br atoms (see Figure \ref{Fig:XPS}a) and to the formation of a mesh of interlinked N-AGNRs with N=3, 6, 9, 12, as reported in Figure \ref{Fig:STM}d. The cyclodehydrogenation is also evident when one observes the C 1s XPS peak that shows a higher C-C component at the expense of the C-H component in comparison to the same signal from PPP chains; see Figure \ref{Fig:XPS}c. It is important to note that on the Ag(111) surface it is not possible to obtain large AGNRs with a defined width, but only a mesh of interlinked nanoribbons. The reason for this significant difference will be discussed in the next section.

\subsection{Formation Energies and Thermodynamics}

\begin{figure*}
	\includegraphics[width=\textwidth]{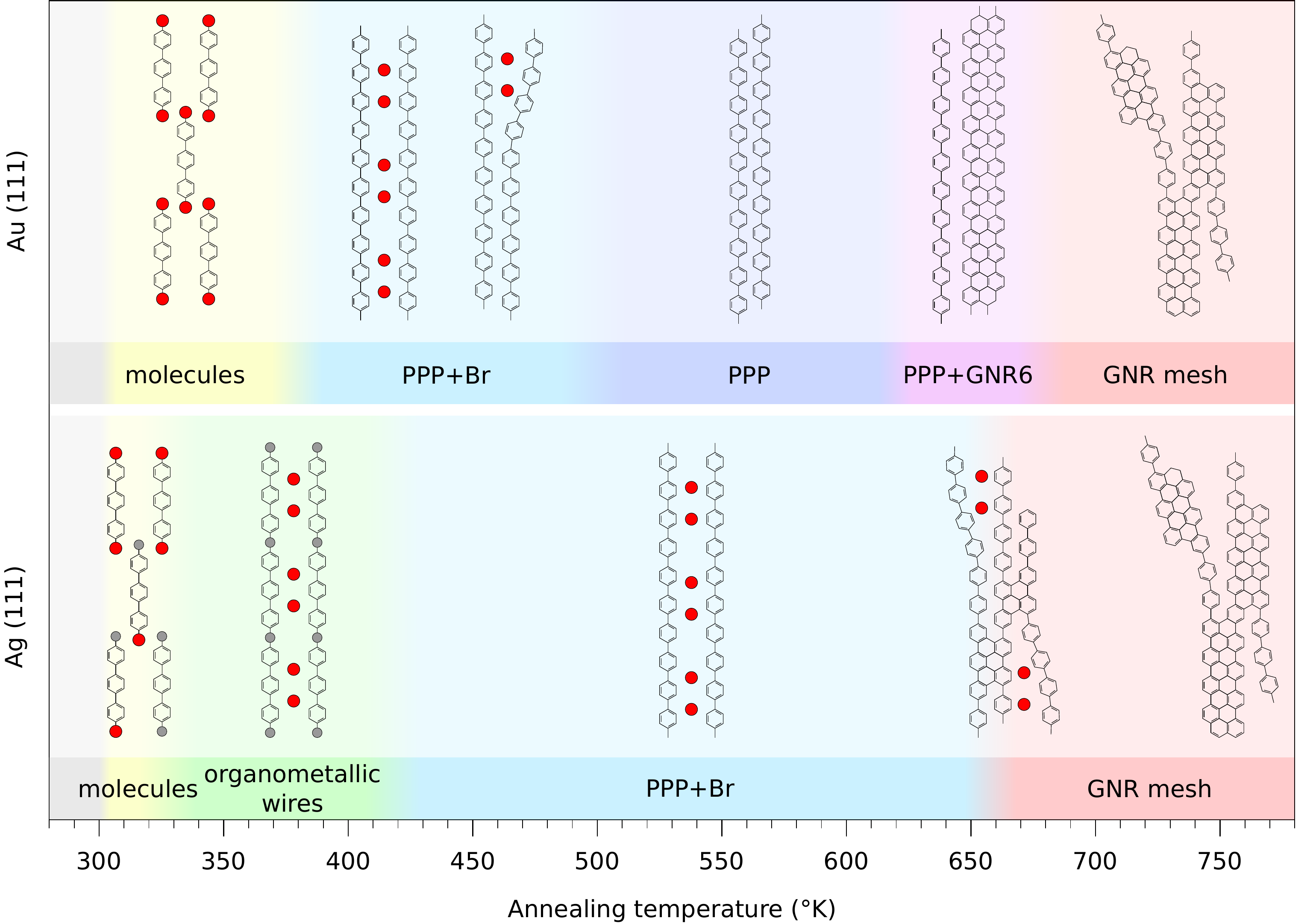}
	\caption{Diagram of the temperature range stability of the different stable nanostructures on Au(111) and Ag(111) surfaces.}
	\label{Fig:temps}
\end{figure*}

The different catalytic activities between Ag(111) and Au(111) surfaces leads to different paths and activation temperatures in the formation of adsorbed nanostructures. In Figure \ref{Fig:temps} we summarize these differences in order to provide a rational representation of the experimental observations. All reported data are the result of a large number of preparations on the two surfaces using a similar protocol, which starts with the depositions of around 1 ML of DBTP molecules followed by a long annealing plateau, for a minimum of 5 hours, at increasing temperatures. At each step the surface has been analyzed by STM, LEED and XPS in order to identify the most stable nanostructures. Transition temperatures are slightly lower than previous reports on similar systems where the growth had been monitored by synchronous XPS measurements during annealing\cite{Vasseur2016} (Figure \ref{Fig:temps}). This discrepancy can be explained considering that the thermodynamic equilibrium is only reached after several hours at fixed temperature and that the relatively fast annealing ramps previously employed do not satisfy this condition.

On the Au(111) surface the chessboard molecular structure is stable at RT with the Br atoms being still attached to the molecules. On the contrary, this structure is only kinetically stabilized on Ag(111) and after some days it transforms into the organometallic chains (OM-PPP), characterized by C-Ag bonds that are still visible after thermal annealing below 400 K. As previously reported, the Au-C bond energy is quite low;\cite{Bejarano2018} indeed, in the case of DBTP molecular deposition on Au(111), the organometallic structures do not appear at any temperature and the system evolves directly from the adsorbed molecules into PPP chains.\cite{Basagni2016}

The probability of formation of the organometallic chains can be inferred by looking at the thermodynamic stability of different adsorbed structures at both the Ag(111) and Au(111) surfaces. The formation energy of PPP chains at the metal surface with intercalated bromine atoms can be evaluated as:

$$
E_{form}^{PPP} = E_{PPP} - E_{DBTP} - N \cdot E_{surf}
$$

where $E_{PPP}$ is the total energy of the adsorbed PPP chain model with intercalated bromines, $E_{DBTP}$ is the energy of the free-standing molecular precursor, $E_{surf}$ is the energy of a slab unit cell and $N$ is the number of slab unit cells in the substrate supercell model. Considering the formation of the organometallic chains, a further term for the chemical potential of the additional metal atoms should be included:

$$
E_{form}^{OM} = {E}_{OM-PPP} - E_{DBTP} - N \cdot E_{surf} - \mu_{M}
$$
where ${E}_{OM-PPP}$ is the total energy of the adsorbed OM-PPP chain model. As  $\mu_{M}$ reference energy we chose the energy of a metal adatom at the metal substrate.

In the case of the silver surface, we obtain a formation energy for the adsorbed organometallic chains of -0.9 eV per carbon ring. This negative value indicates that the adsorption of the DBTP molecules leads to the spontaneous formation of organometallic chains in the presence of a silver adatoms reservoir. Ag-substituted DBTP individual molecules (Figure \ref{Fig:STM}e, j) correspond to an intermediate state toward the complete conversion. The metastable OM-PPP chain state can then transform into the more stable PPP chain configuration for which we obtain a formation energy of -2.0 eV per carbon ring.

In the case of gold, on the contrary, the formation energy for the adsorbed organometallic chains is 0.11 eV per carbon ring while for the PPP chains we obtain -0.7 eV per carbon ring. The positive formation energy of the organometallic chains indicates that they cannot form as a result of the DBTP decomposition and only PPP chains can be assembled at the gold surface.

As reported in Figure \ref{Fig:temps}, on the Au(111) surface a 10 h long annealing around 520 K leads to  nearly complete desorption of the Br atoms,  while the PPP chains remain intact.\cite{Basagni2015,Merino-Diez2017,DeBoni2020,Merino-Diez2018} Increasing the annealing temperature up to 650 K activates the cyclodehydrogenation reaction and two or more neighboring PPP chains, no more separated by Br atoms, can easily close via a zip mechanism forming above all  long 6-AGNRs. On the Ag (111) surface on the contrary, Br atoms desorption and cyclo-dehydrogenation reactions are simultaneously activated at 650 K. As a consequence, the PPP chains can interlink but their full lateral coupling is hindered by the remaining Br atoms. This leads to the formation of a mesh of graphene nanoribbons with different lateral widths.
 
This experimental information indicates that the binding of individual bromine atoms to the metal surface is a key parameter affecting the kinetics of the formation of molecular chains and GNRs. As pointed out in previous works on Au surfaces, the desorption of bromine occurs most probably via the recombination of adsorbed H and Br atoms and the desorption of HBr molecules.\cite{Bronner2015} 

\begin{figure}
	\includegraphics[width=\columnwidth]{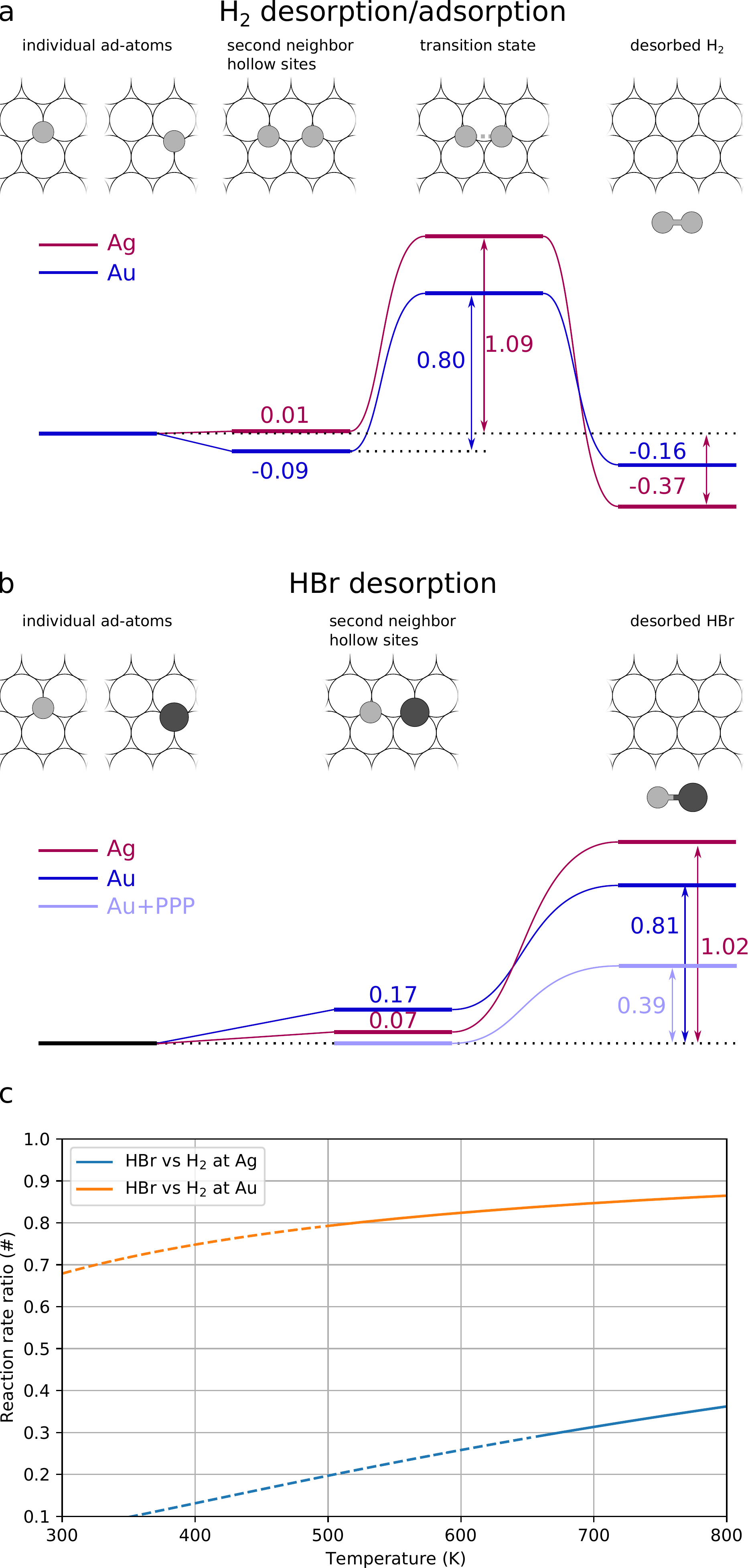}
	\caption{Activation barriers for (a) \ce{H2} and (b) HBr desorption from Ag(111) and Au(111) surfaces. In (b) both the cases of HBr desorption from a clean and PPP covered Au surfaces are presented. (c) Reaction rate ratio for \ce{H2} vs HBr desorption at Ag and Au surfaces.
	}
	\label{Fig:barriers}
\end{figure}

Reaction paths for desorption processes of \ce{H2} and HBr at both Ag (111) and Au (111) surfaces have been derived using the nudged elastic band method combined with the DFT approach. All the obtained energy values have been summarized in Figure \ref{Fig:barriers}. Reaction rates could then be estimated using the Arrhenius equation, but a reasonable guess for the pre-exponential factor would be required. However, the ratio between different reaction rates as a function of temperature can be more easily estimated by assuming that the frequency factor is of the same order of magnitude for all reactions considered here:

$$ 
\frac{r_1}{r_2} = e^\frac{E_2 - E_1}{k_B T}
$$

HBr desorption occurs via the migration of individual H and Br atoms at second neighbor surface hollow sites (which is a metastable state while atoms at first neighbor sites represent a nonstable configuration) and the successive atoms recombination and molecule desorption. For both Ag (111) and Au(111) surfaces, the process is endothermic without intermediate transition states. The desorption activation energy can then be calculated as:

$$
\Delta E = E_{Br @ surf} + E_{H @ surf} - 2 E_{surf} - E_{HBr}
$$

which is the difference between the energy of a free HBr molecule and the chemical potential of H and Br atoms adsorbed at the metal surface. We obtain an activation energy for HBr desorption of 1.02 eV for Ag and 0.81 eV for Au. The HBr formation might be hindered by the concurrent process of direct H-H recombination and desorption. We have therefore calculated the minimum reaction path for H desorption as \ce{H2} on a clean metal surface for which we obtain a transition state with an activation energy of 1.09 eV for Ag and 0.80 eV for Au. The values we obtain for both \ce{H2}  and HBr are slightly higher than previous estimations reported for Au but here we considered thicker Au slabs (14 layers instead of 5 in previous works) which are expected to better represent the electronic structure of the infinite slab.\cite{Bronner2015,Bjork2013} For both surfaces the HBr and \ce{H2}  desorption energies are similar, indicating that the two processes should have close reaction rates (Figure \ref{Fig:barriers}c). However, HBr desorption is a pseudo-first order process (the removal of H from the molecules represents the rate-limiting step; as soon as it is removed it immediately recombines with adsorbed Br) while \ce{H2} desorption is a second order process since it requires the removal and collision of two H atoms.\cite{Mairena2019} 
Therefore, in the case of hydrogen availability, the formation of HBr is kinetically favored over \ce{H2}  desorption until all Br is consumed.

This rationale is usually employed to describe Br desorption and the effective structure of the surface covered by molecular networks is neglected. Considering the presence of PPP chains, hydrogen atoms can only be adsorbed very close to adsorbed Br atoms, at the first or second neighbor hollow site. The activation energy has therefore to be calculated as

$$
\Delta E = E_{Br_{x} + H @ surf + PPP} - E_{Br_{x - 1} surf + PPP} - E_{HBr}
$$

The first term of the sum indicates the energy of the metal surface covered by PPP chains intercalated by Br atoms with one additional H adsorbed. The second term is the energy of the same system where the H and a single Br atom have been desorbed. In this case the activation energy for HBr desorption is strongly lowered to 0.39 eV for Au, thus indicating a strong enhancement of the Br desorption rate when PPP is presented at the surface (Figure \ref{Fig:barriers}b). In the case of the Au (111) surface, it has been discussed that the HBr desorption rate is triggered most likely by the cyclodehydrogenation reaction, which provides the necessary hydrogen.\cite{Bronner2015,Ma2021} Interestingly, contrary to previous observations, in our case the bromine atoms on Au are completely desorbed at 520 K (Figure \ref{Fig:temps}), a lower temperature than that of complete cyclodehydrogenation. This can be explained by the long annealing time employed, about 10 h, which could provide the required H either via partial dehydrogenation or due to the residual hydrogen in the preparation chamber. AuBr ejection has also been proposed as an alternative path for Br desorption\cite{DiGiovannantonio2018,Mairena2019,Bardi1983} but we rule out that this mechanism can play a role at about 500 K due to its very high activation energy, about 2.45 eV as derived by our DFT calculations.

In the case of Ag(111)  partially interlinked chains coexist with individual intercalated Br atoms at the activation temperature for cyclodehydrogenation and the bromine desorption is completed only at around 700 K. This higher desorption temperature is consistent with the calculated larger HBr desorption activation energies discussed above.

\section{Conclusions}
In this work we have presented an experimental and theoretical investigation of the PPP chain growth and subsequent narrow graphene ribbons starting from DBTP molecular precursors deposited at the silver surface. Extended annealing times were employed at each step of the growth to ensure that the thermodynamic equilibrium was reached, and the products were characterized by STM, LEED and XPS. We observed a room temperature debromination of the DBTP molecules and the subsequent substitution of Br atoms with silver atoms. These new organometallic complexes progressively transform into long and well-organized organometallic chains whose structure and ordering have been here described at the atomic level via complementary numerical simulation. While these organometallic complexes spontaneously appear on silver surfaces, they have been very rarely observed on gold. This difference can be explained by looking at the formation energy of organometallic complexes in the presence of metal adatoms reservoirs: while DFT simulations give a negative formation energy for silver organometallic chains, a positive value is obtained in the case of gold.

Mild annealing at about 390 K leads to the decomposition of the organometallic complexes and the formation of PPP chains intercalated by individual Br atoms. Concerning the gold surface, we obtain a flatter conformation of the PPP chains as a result of the strong interaction with the substrate. Br atoms remain at the silver surface up to the cyclodehydrogenation temperature, even when extended annealing times are employed, when they can recombine with hydrogen and desorb as HBr. The dehydrogenative cyclization process coexists then with residual Br atoms intercalating the chains. This situation hinders the formation of well-structured graphene ribbons, leading only to a partial lateral coupling of the PPP chains up to complete Br desorption. This result stresses the importance of the Br desorption process in defining the kinetics of the formation of molecular chains and graphene ribbons in Ullman-like on-surface-synthesis.

In the case of gold instead, complete Br desorption occurs at lower temperatures as a result of the long annealing times. This behavior can be linked with the much lower activation energy obtained for HBr desorption on gold with respect to silver.

\acknowledgements{
This project has received funding from the European Union’s Horizon 2020 research and innovation programme under the Marie Skłodowska-Curie grant agreement No. 842694 and from the University of Padova through grant P-DISC\#09BIRD2019-UNIPD SMOW. Aloisa beamline staff at the Elettra Synchrotron is acknowledged for support during the beamtime.
}

\bibliography{references}

\end{document}